\documentclass[preprint,eqsecnum,aps,nofootinbib]{revtex4}
\usepackage{amsfonts,amsmath,amssymb,amsthm}
\usepackage{latexsym}
\usepackage{bbm,bm}
\usepackage{graphicx}

\newcommand{\ket}[1]{\lvert #1 \rangle}
\newcommand{\bra}[1]{\langle #1 \lvert}
\newcommand{\beq}{\begin{equation}}
\newcommand{\eeq}{\end{equation}}
\newcommand{\beqs}{\begin{eqnarray}}
\newcommand{\eeqs}{\end{eqnarray}}

\begin{document}

\title{Propagation of initial uncertainties to Arthurs-Kelly inequality}

\author{Mi-Ra Hwang$^{1,2}$, Eylee Jung$^{1,2}$, and DaeKil Park$^{1,2,3}$\footnote{corresponding author, dkpark@kyungnam.ac.kr}}

\affiliation{ $^1$ Quanta, Changdong 76, Masan, 51730, Korea     \\
                $^2$Department of Electronic Engineering, Kyungnam University, Changwon
                 631-701, Korea    \\
               $^3$Department of Physics, Kyungnam University, Changwon
                  631-701, Korea  }

\preprint{\bf{KMC-24-01}}
\begin{abstract}

The generalized version of the Arthurs-Kelly inequality is derived when the initial state is a tripartite separable state. When each initial substate obeys the minimal
uncertainty, the generalized version reduces to the well-known inequality, i.e. twice of the Heisenberg uncertainty.  
If the initial probe state is entangled, it is shown that the 
generalized version of the Arthurs-Kelly inequality can be violated. We show the violation explicitly by introducing a special example.

\end{abstract}

\maketitle

\section{Introduction}
The uncertainty principle (UP) is a fundamental concept in quantum physics. The idea of UP was first proposed by Heisenberg\cite{heisenberg1927} by considering the measurement of electron's position by means of scattered light, 
which at the same time leads to a disturbance of the electron's momentum. He estimated the lower bound for the product of the mean errors of position and momentum to be of the order of Planck constant.
The most familiar expression of the UP is Heisenberg-Robertson form\cite{robertson1929};
\begin{equation}
\label{HUP-1}
\Delta_{\hat{x}} \Delta_{\hat{p}} \geq \frac{\hbar}{2}
\end{equation}
where $\left( \Delta_{\hat{A}}^2 \right)_{\psi} \equiv \bra{\psi} \hat{A}^2 \ket{\psi} - \bra{\psi} \hat{A} \ket{\psi}^2$ is a variance computed in the measurement of the observable $\hat{A}$ when the state vector is $\ket{\psi}$.
In the following $\hbar = 1$ will be chosen for simplicity. Also we will drop the quantum state $\ket{\psi}$ when it is evident in a context.

For a long time it has been believed that Eq. (\ref{HUP-1}) states that the position and momentum of a particle cannot be simultaneously measured with arbitrary precision\cite{cohen}.
However, recent study indicates that the inequality (\ref{HUP-1}) is not related to the simultaneous measurement.
This can be understood from the fact that Eq. (\ref{HUP-1}) does not involve the effect of the measurement apparatus. Rather, the interpretation of Eq. (\ref{HUP-1}) is related to a restriction on quantum state preparation\cite{aguilar18,ballentine70,Buss10,ozawa89,appleby98,raymer94}.
This is summarized in Ref.\cite{text}, where the following comment is presented: {\it The correct interpretation of the UP is that if we prepare a large number of quantum systems in identical state, $\ket{\psi}$, and then perform measurements of 
$\hat{x}$ on some of those systems, and of $\hat{p}$ in others, then the standard deviation $\Delta_{\hat{x}}$ times the standard deviation $\Delta_{\hat{p}}$ will satisfy the inequality (\ref{HUP-1})}.

The first quantum-mechanical description of the joint measurement of complementary variables came up with a paper by Arthurs and Kelly (AK) \cite{Ar-Kelly}. 
They generalized the von Neumann measurement process\cite{neumann32}, where a position measurement was considered via a single additional pointer system.
The idea of AK were elaborated further by Arthurs and Goodman in Ref. \cite{ar-88}. More recently, the related topics were discussed in Ref. \cite{busch14}.
For the simultaneous measurement  AK introduced two probes for the measurement process. Thus, the whole system consists of three parties: one of them which is ``3'' in the following is 
a physical particle, whose position and momentum will be jointly measured, and two of them which are ``1'' and ``2'' are the pointer systems. The interaction Hamiltonian is chosen as a form
${\widehat H}_{int} = \kappa \left({\hat x}_3 {\hat p}_1 + {\hat p}_3 {\hat p}_2 \right)$ with large coupling constant $\kappa$.  It is assumed that the joint measurement is performed at time $t = 1 / \kappa$ for simplicity.
Then, AK showed that if the initial state is written as $\Psi_{in} (x_1, x_2, x_3: t=0) = \phi_1(x_1) \phi_2 (x_2) \phi_3 (x_3)$ and each $\phi_j$ is a Gaussian state with minimum uncertainty,
the uncertainty for the meters holds the inequality 
\begin{equation}
\label{Ar-K-1}
\Delta_{\hat{x}_1} (t = 1 / \kappa) \Delta_{\hat{x}_2} (t = 1 / \kappa) \geq 1.
\end{equation}
This is twice of the Heisenberg uncertainty given in Eq. (\ref{HUP-1}). The experimental test of the joint measurement was also discussed in Ref. \cite{exp1,exp2,exp3}.

In this letter we examine the dependence of the initial state in the AK inequality (\ref{Ar-K-1}). In particular, we will examine the effect of the initial entangled probe state in detail.
For the purpose we choose the initial state as  $\Psi_{in} (x_1, x_2, x_3: t=0) = \psi (x_1, x_2) \phi_3 (x_3)$, where $\psi$ and $\phi$ are arbitrary normalized wave function. 
The time evolution of quantum state will be determined by deriving the Feynman propagator\cite{feynman,kleinert} of the AK system explicitly. In next section the main results of this letter is summarized. In section III the Feynman propagator of the AK system is explicitly derived. 
In section IV the main results introduced in section II are derived. Also, in order to discuss the effect of the entangled probe states, we introduce a special example in this section. In section V a brief conclusion is given.

\section{Main results}
The final and main results of this letter can be summarized as following. At time $t = 1 / \kappa$ the variances $\Delta_{\hat{x}_1}^2 \left(t = 1 / \kappa \right)$ and $\Delta_{\hat{x}_2}^2 \left(t = 1 /\kappa \right)$ are expressed in terms of the variances at $t=0$ as following:
\begin{eqnarray}
\label{main-1}
&&  \Delta_{\hat{x}_1}^2 \left(t = 1 / \kappa \right) = \Delta_{\hat{x}_1}^2  + \Delta_{\hat{x}_3}^2  + \frac{1}{4} \Delta_{\hat{p}_2}^2  + \alpha          \\   \nonumber
&&  \Delta_{\hat{x}_2}^2 \left(t = 1 / \kappa \right) = \Delta_{\hat{x}_2}^2  + \Delta_{\hat{p}_3}^2  + \frac{1}{4} \Delta_{\hat{p}_1}^2  - \beta
\end{eqnarray}
where 
\begin{equation}
\label{main-2}
\alpha = \langle \hat{x}_1 \hat{p}_2 \rangle  - \langle \hat{x}_1 \rangle  \langle \hat{p}_2 \rangle     \hspace{1.0cm}   \beta = \langle \hat{x}_2 \hat{p}_1 \rangle  - \langle \hat{x}_2 \rangle  \langle \hat{p}_1 \rangle .
\end{equation}
In the following any variances and means without explicit time are assumed to be computed at $t = 0$. Therefore, Eq. (\ref{main-1}) shows the propagation of uncertainties from $t = 0$ to $t = 1 / \kappa$ for large $\kappa$. 

If the initial probe states is uncorrelated, i.e. $\psi(x_1, x_2) = \phi_1 (x_1) \phi_2 (x_2)$, then $\alpha = \beta = 0$, and the inequality (\ref{Ar-K-1})
is generalized as a form:
\begin{equation}
\label{final-1}
\Delta_{\hat{x}_1} (t = 1 / \kappa) \Delta_{\hat{x}_2} (t = 1 / \kappa) \geq \Gamma = \Gamma_1 + \Gamma_2 + \Gamma_3
\end{equation}
where $\Gamma_j$ is a contribution of each probe states and system as a form
\begin{equation}
\label{final-2}
\Gamma_1 = \frac{1}{2} \Delta_{\hat{x}_1}  \Delta_{\hat{p}_1}      \hspace{.5cm} \Gamma_2 = \frac{1}{2} \Delta_{\hat{x}_2}  \Delta_{\hat{p}_2}   \hspace{.5cm} \Gamma_3 =  \Delta_{\hat{x}_3}  \Delta_{\hat{p}_3}.
\end{equation}
If each state obeys the minimal uncertainty, it is obvious that Eq. (\ref{Ar-K-1}) is reproduced. 
If initial probe states are correlated with each other, it is shown that the generalized AK inequality (\ref{final-1}) can be violated because of nonzero of $\alpha$ and $\beta$. 
This fact will be explicitly illustrated by choosing a general entangled Gaussian probe state.

\section{Feynman Propagator}
In order to derive the Feynman propagator we need to introduce the kinetic Hamiltonian as well as the interaction Hamiltonian. Since we will take the large coupling constant, it is free to choose any kinetic Hamiltonian 
because its contribution to the uncertainties is negligible. Thus, we will take the simplest one, the free particle Hamiltonian.
Thus the total Hamiltonian is 
\begin{equation}
\label{hamil-1}
\widehat{H} = \widehat{H}_{free} + \widehat{H}_{int}
\end{equation}
where
\begin{equation}
\label{hamil-2}
\widehat{H}_{free} = \frac{\hat{p}_1^2}{2 m_1} + \frac{\hat{p}_2^2}{2 m_2} +  \frac{\hat{p}_3^2}{2 m_3} 
\hspace{1.0cm} \widehat{H}_{int} = \kappa \left(\hat{x}_3 \hat{p}_1 + \hat{p}_3 \hat{p}_2 \right).
\end{equation}

Now, we consider the classical mechanics with total  Hamiltonian\footnote{Since we are considering the classical mechanics and path-integral quantum mechanics in this section, the hat ,which denotes operator, will be dropped.} $H$.
From the total Hamiltonian the classical Hamilton equations $\dot{x}_j = \frac{\partial H}{\partial p_j}$ and $\dot{p}_j = -\frac{\partial H}{\partial x_j}$ can be easily derived.  
Eliminating the conjugate momenta $p_j$ from the Hamilton equations, one can derive the classical equations of motion in the form:
\begin{eqnarray}
\label{eqofmotion1}
&& \ddot{x}_1 - \kappa \dot{x}_3 = 0  \hspace{1.0cm} \ddot{x}_2 - \kappa m_3 \ddot{x}_3 = 0  \hspace{.5cm}     \\   \nonumber
&&\frac{m_3}{b} \left(m_2 \kappa \ddot{x}_2 - \ddot{x}_3 \right) = - m_1 \kappa (\dot{x}_1 - \kappa x_3)
\end{eqnarray}
where 
\begin{equation}
\label{hamil-10}
b = m_2 m_3 \kappa^2 - 1.
\end{equation}

For the derivation of the Feynman propagator we need a action, from which the classical equations of motion (\ref{eqofmotion1}) are derived from the least action principle.
The action can be explicitly derived by a trial and error in the form:
\begin{equation}
\label{action1}
S = \int_0^t d t L (\dot{x}_j, x_j)
\end{equation}
where the Lagrangian $L (x_j, \dot{x}_j)$ is 
\begin{equation}
\label{lagrangian}
L = \frac{m_1}{2}  (\dot{x}_1 - \kappa x_3)^2 - \frac{1}{2 b} \left( m_2 \dot{x}_2^2 + m_3 \dot{x}_3^2 - 2 \kappa m_2 m_3 \dot{x}_2 \dot{x}_3 \right).
\end{equation}

Generally, the Feynman propagator can computed by a path-integral
\begin{equation}
\label{path-1}
K[x_j (t) : x_j(0): t] = \int D x_1 D x_2 D x_3 e^{ i S}.
\end{equation}
Since, however, the action in Eq. (\ref{action1}) is quadratic, we do not need to compute the path-integral explicitly. It is expressed in terms of the classical action $S_{cl}$ and the prefactor $F(t)$  in the form
\begin{equation}
\label{path-2}
K[x_j (t): x_j(0): t] = F(t) e^{ i S_{cl}}.
\end{equation}

In order to compute the classical action we should solve Eq. (\ref{eqofmotion1}) explicitly. Inserting the classical solutions into the action (\ref{action1}) one can derive the classical action as following:
\begin{eqnarray}
\label{cl_action1}
&& S_{cl} = \frac{1}{2 a(t) b t} \Bigg[ 12 m_1 m_3 b z_1^2 - m_2 a(t) z_2^2 - m_3 a(t) z_-^2 + 3 m_1 m_3 \kappa^2 t^2 b z_+^2         \\   \nonumber
&&   \hspace{4.0cm}   - 12 m_1 m_3 \kappa t b z_1 z_+ + 2 m_2 m_3 \kappa a(t) z_2 z_- \Bigg].
\end{eqnarray}
where $ a(t) = 12 m_3 + m_1 \kappa^2 t^2$ and 
\begin{equation}
\label{eqofmotion3}
z_1 = Q_1 - q_1    \hspace{.5cm}  z_2 = Q_2 - q_2  \hspace{.5cm} z_{\pm} = Q_3 \pm q_3 .
\end{equation}
When deriving the classical action we used the boundary conditions as $x_j (0) = q_j$ and $x_j (T) = Q_j$.
When $\kappa = 0$, $S_{cl}$ reduces to 
\begin{equation}
\label{cl_action2}
S_{cl} = \frac{m_1 z_1^2 + m_2 z_2^2 + m_3 z_-^2}{2 t}.
\end{equation}
This is just the classical action for three free particle systems.

Now, let us consider the prefactor $F(t)$. In the following we will compute $F(t)$ explicitly without using the path-integral.
First we note that if $\kappa = 0$, $F(t)$ should become
\begin{equation}
\label{prefactor1}
F(t) = \sqrt{\frac{m_1}{2 \pi i t}} \sqrt{\frac{m_2}{2 \pi i t}} \sqrt{\frac{m_3}{2 \pi i t}} = \sqrt{\frac{i m_1 m_2 m_3}{8 \pi^3 t^3}}.
\end{equation}
In order to compute $F(t)$ we use the identity\cite{feynman,kleinert}
\begin{eqnarray}
\label{prefactor2}
&&  K[Q_1, Q_2, Q_3: q_1, q_2,  q_3: t]                   \\    \nonumber
&&= \int dx_1 dx_2 dx_3 K[Q_1, Q_2, Q_3: x_1, x_2, x_3: t_1] K[x_1, x_2, x_3: q_1, q_2, q_3: t - t_1],
\end{eqnarray}
which yields
\begin{equation}
\label{prefactor3}
F(t) = \pi^{3/2} F(t_1) F(t - t_1) \sqrt{\frac{2 i b t_1^3 (t - t_1)^3 a(t_1) a (t - t_1)}{3 m_1 m_2 m_3^2 t^3 a(t)}}.
\end{equation}
Solving Eq. (\ref{prefactor3}), one can show 
\begin{equation}
\label{prefactor4}
F(t) = \sqrt{\frac{3 m_1 m_2 m_3^2}{2 \pi^3 i b t^3 a(t)}}.
\end{equation}
When $\kappa = 0$, $a = 12 m_3$ and $b = -1$, which makes Eq. (\ref{prefactor4}) coincide with Eq. (\ref{prefactor1}) exactly.
Therefore, the Feynman propagator or Kernel for the system (\ref{hamil-1}) is 
\begin{equation}
\label{feynman1} 
K[ Q_1, Q_2, Q_3: q_1, q_2, q_3: t] =  \sqrt{\frac{3 m_1 m_2 m_3^2}{2 \pi^3 i b t^3 a(t)}} e^{i S_{cl}}
\end{equation}
where $S_{cl}$ is explicitly given in Eq. (\ref{cl_action1}). 
One can derive the Feynman propagator (\ref{feynman1}) differently by making use of the time evolution operator given in Ref. \cite{free_evolution}, 
where the effect of free particle evolution is examined in AK inequality without adopting the large $\kappa$ assumption.

\section{Derivation of Main results}

Since the Feynman propagator is a coordinate representation of the time-evolution operator, it is possible to derive the wave function at $t=T$ from the 
wave function at $t = 0$ as follows:
\begin{equation}
\label{evolution-1}
\Psi(x_1, x_2, x_3: T) = \int dq_1 dq_2 dq_3 K[x_1, x_2, x_3: q_1, q_2, q_3:T] \psi(q_1, q_2) \phi_3 (q_3).
\end{equation}
It is convenient to express the Feynman propagator in a form:
\begin{equation}
\label{feynman2}
 K[x_1, x_2, x_3: q_1, q_2, q_3:T] = \sqrt{\frac{3 m_1 m_2 m_3^2}{2 \pi^3 i a b T^3}} e^{i S_{cl}}
 \end{equation}
 where $a = a(T) = 12 m_3 + m_1 \kappa^2 T^2$ and $S_{cl}$ is 
 \begin{eqnarray}
 \label{cl_action3}
 &&S_{cl} =  \frac{f(x_1, x_2, x_3)}{2 a b T} + A_1 q_1^2 + A_2 q_2^2 + A_3 q_3^2 + 2 A_{13} q_1 q_3        \\    \nonumber
 && \hspace{3.0cm}+ 2 A_{23} q_2 q_3 + B_1 q_1 + B_2 q_2 + B_3 q_3.
 \end{eqnarray}
 In Eq. (\ref{cl_action3}) $f(x_1, x_2, x_3)$ is 
 \begin{eqnarray}
 \label{cl_action3-boso1}
&&  f(x_1, x_2, x_3) =  12 m_1 m_3 b x_1^2 - m_2 a x_2^2 - (m_3 a - 3 m_1 m_3 \kappa^2 T^2 b) x_3^2       \\     \nonumber
&&   \hspace{3.0cm}  - 12 m_1 m_3 \kappa T b x_1 x_3 + 2 m_2 m_3 \kappa a x_2 x_3
 \end{eqnarray}
 and the coefficients are
 \begin{eqnarray}
 \label{ac_action3-boso2}
&&  A_1 = \frac{6 m_1 m_3}{a T}   \hspace{.5cm}  A_2 = - \frac{m_2}{2 b T}      \hspace{.5cm}  A_3 = \frac{m_3 (3 m_1 \kappa^2 T^2 b - a)}{2 a b T}   \hspace{.5cm}  A_{13} = \frac{3 m_1 m_3 \kappa}{a}     \\    \nonumber
&& A_{23} = \frac{m_2 m_3 \kappa}{2 b T}   \hspace{.5cm}  B_1 = \frac{6 m_1 m_3 (\kappa T x_3 - 2 x_1)}{a T}     \hspace{.5cm}  B_2 = \frac{m_2 (x_2 - m_3 \kappa x_3)}{b T}                                               \\    \nonumber
&& B_3 = \frac{m_3}{a b T} \bigg[ (a + 3 m_1 \kappa^2 T^2 b) x_3 - 6 m_1 \kappa T b x_1 - m_2 \kappa a x_2 \bigg].
 \end{eqnarray}
It is worthwhile noting that among all coefficients only $B_j \hspace{.1cm} (j=1, 2, 3)$ depend on $x_j$. 
Therefore,  $\Psi(x_1, x_2, x_3:T)$ can be written as 
\begin{eqnarray}
\label{g-AK-2}
&&\Psi(x_1, x_2, x_3:T)  = \sqrt{\frac{3 m_1 m_2 m_3^2}{2 \pi^3 i a  b T^3}} \exp \left[ \frac{i}{2 a b T} f(x_1, x_2, x_3) \right]    \\    \nonumber
&&\times \int dq_1 dq_2 dq_3 \exp \bigg[ i A_1 q_1^2 + i A_2 q_2^2 + i A_3 q_3^2 + 2 i A_{13} q_1 q_3 + 2 i A_{23} q_2 q_3                    \\    \nonumber
&&\hspace{6.0cm}      + i B_1 q_1 + i B_2 q_2 + i B_3 q_3 \bigg]  \psi(q_1, q_2) \phi_3 (q_3).
\end{eqnarray}

From this stage we will  take $T = 1 / \kappa$ for simplicity. Furthermore, we will take  large $\kappa$ limit subsequently.
Therefore, the following equations are valid only in this limit.
In this case the coefficients becomes $a(T) \equiv a = m_1 + 12 m_3$ and $b \approx m_2 m_3 \kappa^2$. The other coefficients given in Eq. (\ref{ac_action3-boso2}) should be replaced with their dominant terms in this limit.
%
Then, the normalization condition of  $\Psi(x_1, x_2, x_3:T) $ can be proved by noting
\begin{equation}
\label{g-AK-5}
\int dx_1 dx_2 dx_3 e^{i B_1 (q_1 - q_1') + i B_2 (q_2 - q_2') + i B_3 (q_3 - q_3')} = \frac{(2 \pi)^3 a}{12 m_1 m_3 \kappa} \delta(\theta_1) \delta (\theta_2) \delta(\theta_3)
\end{equation}
where
\begin{eqnarray}
\label{g-AK-6}
&&\theta_1 = (q_1 - q_1' )+ \frac{1}{2} (q_3 - q_3')  \hspace{1.0cm} \theta_2 = \frac{1}{m_3 \kappa} (q_2 - q_2') - (q_3 - q_3')        \\   \nonumber
&&\theta_3 = \frac{6 m_1 m_3 \kappa}{a} (q_1 - q_1') - (q_2 - q_2') + \frac{3 m_1 m_3 \kappa}{a} (q_3 - q_3').
\end{eqnarray}
When deriving Eq. (\ref{g-AK-5}) we used $\delta(a x) = \delta(x) / |a|$. 
Furthermore, one can show 
\begin{equation}
\label{g-AK-7}
\delta(\theta_1) \delta (\theta_2) \delta(\theta_3) = \delta(q_1 - q_1') \delta(q_2 - q_2') \delta(q_3 - q_3'),
\end{equation}
which is proved in appendix A. Using Eq. (\ref{g-AK-7}) it is easy to show $\int dx_1 dx_2 dx_3 |\Psi(x_1, x_2, x_3:T)|^2 = \int dq_1 dq_2 |\psi (q_1, q_2)|^2 \int dq_3 |\phi(q_3)|^2 = 1$. 

\subsection{calculation of $ \Delta_{\hat{x}_1}^2  (t = 1 / \kappa)$}

For the derivation of $\langle \hat{x}_1 \rangle  (t = 1 / \kappa)$ we need a following formula
\begin{eqnarray}
\label{g-AK-9}
&&\hspace{2.0cm} \int dq_1 dq_2 dq_3 \delta'(\theta_1) \delta(\theta_2) \delta (\theta_3) g(q_1, q_2, q_3)         \\    \nonumber 
&&= - \left[ \left(1 - \frac{3 m_1}{a} \right) \frac{\partial}{\partial q_1'} + \frac{6 m_1 m_3 \kappa}{a} \frac{\partial}{\partial q_2'} + \frac{6 m_1}{a} \frac{\partial}{\partial q_3'} \right] g(q_1', q_2, q_3').
\end{eqnarray}    
where the prime in delta function means the derivative with respect to the argument.       
This is proved in appendix A. By making use of Eq. (\ref{g-AK-9})  it is straightforward to show 
\begin{eqnarray}
\label{g-AK-10}
&&\langle \hat{x}_1 \rangle  \left(t = 1 / \kappa \right)  = \frac{a}{12 m_1 m_3 \kappa} \left(1 - \frac{3 m_1}{a} \right) \langle \hat{p}_1 \rangle  + \frac{1}{2} \langle \hat{p}_2 \rangle  + \frac{1}{2 m_3 \kappa} \langle \hat{p}_3 \rangle         \\    \nonumber
&&\hspace{.5cm} +  \left[ \frac{a}{6 m_1 m_3 \kappa} \left( 1 - \frac{3 m_1}{a} \right) A_1 + \frac{1}{ m_3 \kappa} A_{13} \right] \langle \hat{x}_1 \rangle  + \left( A_2 + \frac{1}{2 m_3 \kappa} A_{23} \right) \langle \hat{x}_2 \rangle   \\   \nonumber
&& \hspace{1.0cm}+ \left[ \frac{a}{6 m_1 m_3 \kappa} \left( 1 - \frac{3 m_1}{a} \right) A_{13} + A_{23} + \frac{1}{m_3 \kappa} A_3 \right] \langle \hat{x}_3 \rangle.
\end{eqnarray}
Taking large $\kappa$ limit it is easy to  derive
\begin{equation}
\label{g-AK-11}
\langle \hat{x}_1 \rangle  \left(t = 1 / \kappa \right) \approx \langle \hat{x}_1 \rangle  + \langle \hat{x}_3 \rangle  + \frac{1}{2} \langle \hat{p}_2 \rangle.
\end{equation}

 Appendix A also proves
\begin{eqnarray}
\label{g-AK-13}
&&\hspace{2.0cm} \int dq_1 dq_2 dq_3 \delta''(\theta_1) \delta(\theta_2) \delta (\theta_3) g(q_1, q_2, q_3)         \\    \nonumber 
&&=  \left[ \left(1 - \frac{3 m_1}{a} \right) \frac{\partial}{\partial q_1'} + \frac{6 m_1 m_3 \kappa}{a} \frac{\partial}{\partial q_2'} + \frac{6 m_1}{a} \frac{\partial}{\partial q_3'} \right]^2 g(q_1', q_2, q_3').
\end{eqnarray}   
 Then, it is straightforward to show      
\begin{eqnarray}
\label{g-AK-14}
&& \langle \hat{x}_1^2 \rangle  \left(t = 1 / \kappa \right)                            
= \langle \hat{x}_1^2 \rangle  + \langle \hat{x}_3^2 \rangle  +2 \langle \hat{x}_1 \rangle  \langle \hat{x}_3 \rangle  +  \frac{a - 3 m_1}{6 m_1 m_3 \kappa} \left[ \langle \hat{x}_1 \hat{p}_1 \rangle  + \langle \hat{x}_3 \rangle  \langle \hat{p}_1 \rangle  \right]    \\    \nonumber
&& \hspace{1.0cm} + \langle \hat{x}_1 \hat{p}_2 \rangle  + \langle \hat{p}_2 \rangle  \langle \hat{x}_3 \rangle  + \frac{1}{m_3 \kappa} \left[ \langle \hat{x}_3 \hat{p}_3 \rangle  + \langle \hat{x}_1 \rangle  \langle \hat{p}_3 \rangle  \right]                   
+ \frac{a - 3 m_1}{12 m_1 m_3 \kappa} \langle \hat{p}_1 \hat{p}_2 \rangle                                                                                                   \\   \nonumber
&&+ \frac{a - 3 m_1}{12 m_1 m_3^2 \kappa^2} \langle \hat{p}_1 \rangle  \langle \hat{p}_3 \rangle  + \frac{1}{2 m_3 \kappa} \langle \hat{p}_2 \rangle  \langle \hat{p}_3 \rangle     
 +\left( \frac{a - 3 m_1}{12 m_1 m_3 \kappa} \right)^2 \langle \hat{p}_1^2 \rangle  + \frac{1}{4} \langle \hat{p}_2^2 \rangle  + \frac{1}{4 m_3^2 \kappa^2} \langle \hat{p}_3^2 \rangle                               \\    \nonumber
 && \approx \langle \hat{x}_1^2 \rangle  + \langle \hat{x}_3^2 \rangle  + 2 \langle \hat{x}_1 \rangle  \langle \hat{x}_3 \rangle  + \langle \hat{x}_1 \hat{p}_2 \rangle  + \langle \hat{p}_2 \rangle  \langle \hat{x}_3 \rangle  + \frac{1}{4} \langle \hat{p}_2^2 \rangle.
 \end{eqnarray}
The final expression in the right hand of Eq. (\ref{g-AK-14}) is derived by taking large $\kappa$ limit.
 Therefore $\Delta_{\hat{x}_1}^2 (t = 1 / \kappa) \equiv \langle \hat{x}_1^2 \rangle (t = 1 / \kappa) - \langle \hat{x}_1 \rangle^2 (t = 1 / \kappa)$ can be computed from Eqs. (\ref{g-AK-11}) and (\ref{g-AK-14}), which results in first equation of Eq. (\ref{main-1}).
 
 \subsection{calculation of $\Delta_{\hat{x}_2}^2 (t = 1 / \kappa)$}
 Applying similar way one can compute $\langle \hat{x}_2 \rangle (t = 1 / \kappa)$ and $\langle \hat{x}_2^2 \rangle (t = 1 / \kappa)$, whose explicit expressions are 
 \begin{eqnarray}
 \label{g-AK-16}
 && \langle \hat{x}_2 \rangle (t = 1 / \kappa) \approx \langle \hat{x}_2 \rangle  + \langle \hat{p}_3 \rangle  - \frac{1}{2} \langle \hat{p}_1 \rangle                      \\   \nonumber
 && \langle \hat{x}_2^2 \rangle (t = 1 / \kappa) \approx \langle \hat{x}_2^2 \rangle  + \langle \hat{p}_3^2 \rangle  + 2 \langle \hat{x}_2 \rangle  \langle \hat{p}_3 \rangle  - \langle \hat{x}_2 \hat{p}_1 \rangle  - \langle \hat{p}_1 \rangle  \langle \hat{p}_3 \rangle  
 + \frac{1}{4} \langle \hat{p}_1^2 \rangle.
 \end{eqnarray}
 Thus, $\langle \hat{x}_2 \rangle (t = 1 / \kappa)$ and $\langle \hat{x}_2^2 \rangle (t = 1 / \kappa)$ are derived from $\langle \hat{x}_1 \rangle (t = 1 / \kappa)$ and $\langle \hat{x}_1^2 \rangle (t = 1 / \kappa)$ by changing $\hat{x}_1 \rightarrow \hat{x}_2$, $\hat{x}_3 \rightarrow \hat{p}_3$, 
 and $\hat{p}_2 \rightarrow -\hat{p}_1$.
 Using this rule one can derive the second equation of Eq. (\ref{main-1}).

 \subsection{Generalized Arthurs-Kelly inequality : case for separable probe state}
 
 If the probe states is $\psi(x_1, x_2) = \phi_1 (x_1) \phi_2 (x_2)$, it is easy to prove $\alpha = \beta = 0$. 
Let us define $K_j \equiv \Delta_{\hat{x}_j}^2  \Delta_{\hat{p}_j}^2 $. Then, it is easy to show
\begin{eqnarray}
\label{g-AK-18}
&&\Delta_{\hat{x}_1}^2 \left(t = 1 / \kappa \right) \Delta_{\hat{x}_2}^2 \left(T = 1 / \kappa \right)                           \\     \nonumber
&&= \frac{1}{4} (K_1 + K_2) + K_3 + \frac{K_1 K_2}{x} + \frac{K_2 K_3}{y} + K_1 \frac{y}{x} + \frac{K_3}{4} \frac{x}{y} + \frac{x}{16} + \frac{y}{4}
\end{eqnarray}
where $x = \Delta_{\hat{p}_1}^2  \Delta_{\hat{p}_2}^2$ and  $y = \Delta_{\hat{p}_2}^2  \Delta_{\hat{p}_3}^2 $. This is minimized if the equations 
\begin{eqnarray}
\label{g-AK-19}
&&- \frac{K_1 K_2}{x^2} - K_1 \frac{y}{x^2} + \frac{K_3}{4} \frac{1}{y} + \frac{1}{16} = 0                    \\    \nonumber
&&- \frac{K_2 K_3}{y^2} + \frac{K_1}{x} - \frac{K_3}{4} \frac{x}{y^2} + \frac{1}{4} = 0
\end{eqnarray}
hold. Eq. (\ref{g-AK-19}) is solved by $x = 4 \sqrt{K_1 K_2}$ and $y = 2 \sqrt{K_2 K_3}$. Inserting $x$ and $y$ into Eq. (\ref{g-AK-18}) one can derive the main result (\ref{final-1}). 

 \subsection{Case for correlated probe state}
 In this case $\alpha$ and $\beta$ are generally nonzero and nonnegative.
From Eq. (\ref{main-1}) One can show that $\Delta_{\hat{x}_1}^2 (t = 1 / \kappa) \Delta_{\hat{x}_2}^2 (t = 1 / \kappa)$ can be written as 
\begin{eqnarray}
\label{g-AK-20}
&&\Delta_{\hat{x}_1}^2 \left(t = 1 / \kappa \right) \Delta_{\hat{x}_2}^2 \left(t = 1 / \kappa \right)                           \\     \nonumber
&=& \frac{1}{4} (K_1 + K_2) + K_3 + \frac{K_1 K_2}{x} + \frac{K_2 K_3}{y} + K_1 \frac{y}{x} + \frac{K_3}{4} \frac{x}{y} + \frac{x}{16} + \frac{y}{4} - \alpha \beta     \\    \nonumber
&& + \alpha \left[ K_2 \sqrt{\frac{z}{x y}} + \sqrt{\frac{y z}{x}} + \frac{1}{4} \sqrt{\frac{x z}{y}} \right] - \beta \left[ K_1 \sqrt{\frac{y}{x z}} + K_3 \sqrt{\frac{x}{y z}} + \frac{1}{4} \sqrt{\frac{x y}{z}} \right]
\end{eqnarray}
where $z = \Delta_{\hat{p}_3}^2  \Delta_{\hat{p}_1}^2$. This does not seem to have physically relevant lower bound because equation $\frac{\partial}{\partial z} \Delta_{\hat{x}_1}^2 \left(t = 1 / \kappa \right) \Delta_{\hat{x}_2}^2 \left(t = 1 / \kappa \right)  = 0$
does not provide the positive solution for $z$. Of course, one can make a lower bound by removing $\alpha \left[ K_2 \sqrt{\frac{z}{x y}} + \sqrt{\frac{y z}{x}} + \frac{1}{4} \sqrt{\frac{x z}{y}} \right]$ term in Eq. (\ref{g-AK-20}). However, this does not seem to be the tight lower bound.

Eq. (\ref{g-AK-20}) implies the violation of the generalized AK inequality (\ref{final-1}) or more strongly the original AK inequality (\ref{Ar-K-1}) depending on $\alpha$ and $\beta$ . In order to confirm this fact we consider the special example in the following, where the general Gaussian probe state is chosen.
In this special example we find the violation of (\ref{final-1}), but unfortunately, the violation of (\ref{Ar-K-1}) is not found.


\subsection{special example}
Let us choose the initial correlated probe state as a following general Gaussian state
\begin{equation}
\label{example-1}
\psi (x_1, x_2) = {\cal N} \exp \left[ -\frac{A}{2} x_1^2 - \frac{B}{2} x_2^2 + C x_1 x_2 + D_1 x_1 + D_2 x_2 \right]
\end{equation}
where $A$, $B$, $C$, $D_1$, and $D_2$ are complex numbers. Of course, ${\cal N}$ is a normalization constant given by
\begin{equation}
\label{example-2}
|{\cal N}|^2 = \frac{\sqrt{A_R B_R - C_R^2}}{\pi} \exp \left[ - \frac{A_R D_{2, R}^2 + B_R D_{1, R}^2 + 2 D_{1, R} D_{2, R} C_R}{A_R B_R - C_R^2} \right]
\end{equation}
where $z_R = Re z$ and $z_I = Im z$. After some calculation on can show
\begin{eqnarray}
\label{example-3}
&&\Delta_{\hat{x}_1}^2  = \frac{1}{2} \frac{B_R}{A_R B_R - C_R^2}    \hspace{1.0cm} \Delta_{\hat{x}_2}^2  = \frac{1}{2} \frac{A_R}{A_R B_R - C_R^2}                              \\    \nonumber
&&\Delta_{\hat{p}_1}^2  = \frac{A_R}{2} + \frac{A_R C_I^2 + A_I^2 B_R - 2 A_I C_R C_I}{2 (A_R B_R - C_R^2)}                       \\    \nonumber
&&\Delta_{\hat{p}_2}^2  = \frac{B_R}{2} + \frac{A_R B_I^2 + B_R C_I^2 - 2 B_I C_R C_I}{2 (A_R B_R - C_R^2)}                       \\    \nonumber
&&\alpha \equiv \langle \hat{x}_1 \hat{p}_2 \rangle  - \langle \hat{x}_1 \rangle  \langle \hat{p}_2 \rangle  = \frac{B_R C_I - B_I C_R}{2 (A_R B_R - C_R^2)}                  \\    \nonumber
&&\beta \equiv \langle \hat{x}_2 \hat{p}_1 \rangle  - \langle \hat{x}_2 \rangle  \langle \hat{p}_1 \rangle  = \frac{A_R C_I - A_I C_R}{2 (A_R B_R - C_R^2)}.    
\end{eqnarray}
Thus, the imaginary parts of $A$, $B$, and $C$ make $\alpha$ and $\beta$ to be nonzero.

\begin{figure}[ht!]
\begin{center}
\includegraphics[height=6.0cm]{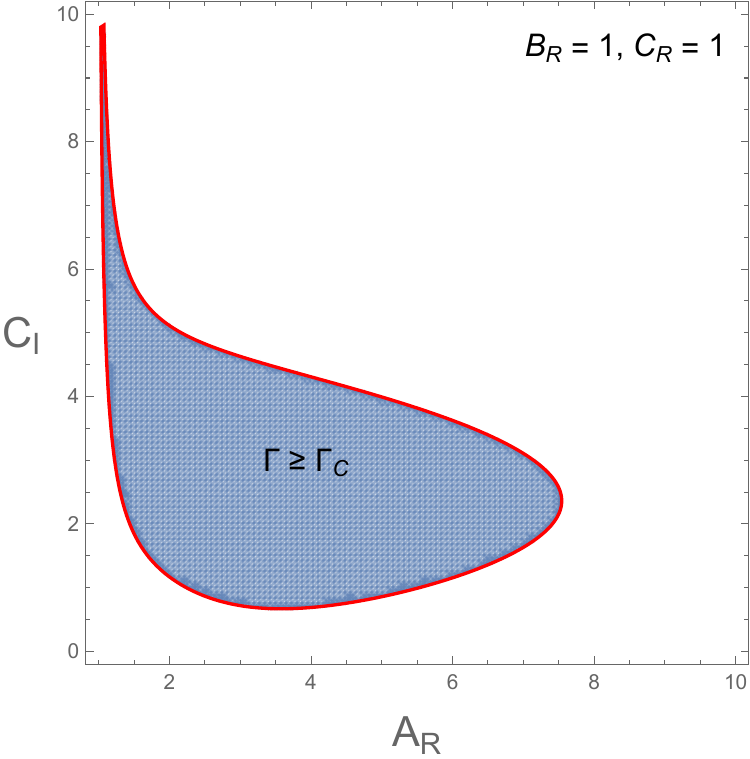} 
\includegraphics[height=6.0cm]{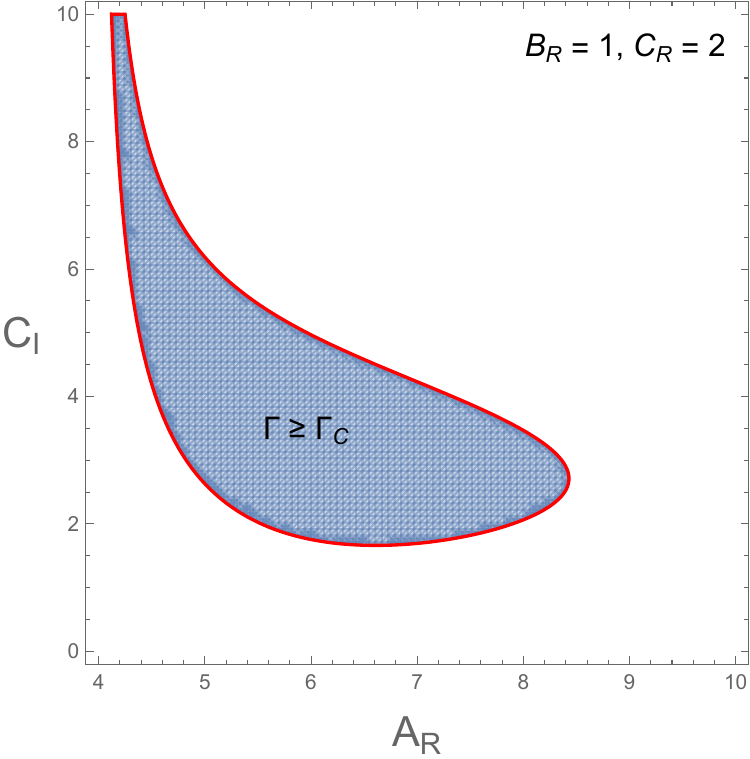}

\caption[fig1]{(Color online) We compare the lower bound $\Gamma$ for the uncorrelated case in Eq. (\ref{example-6}) and $\Gamma_C= \Delta_{x_1} (T = 1 / \kappa) \Delta_{x_2} (T = 1/\kappa)$ for correlated case given in  (\ref{example-7}) when (a) $B_R = C_R = 1$ and (b) $B_R = 1$ and $C_R = 2$.  
The red boundary regions in (a) and (b) are regions where $\Gamma_C \leq \Gamma$. In this sense the generalized AK inequality (\ref{final-1}) is violated in this region. Unfortunately, we cannot find the violation of the original AK inequality expressed in Eq. (\ref{Ar-K-1}) in this example. }
\end{center}
\end{figure}

Now, for simplicity, we set 
\begin{equation}
\label{example-4}
A_I = \frac{C_R C_I}{B_R}   \hspace{1.0cm} B_I = \frac{C_R C_I}{A_R}.
\end{equation}
In this case we get 
\begin{equation}
\label{example-5}
\Delta_{\hat{x}_1}^2  \Delta_{\hat{p}_1}^2  = \Delta_{\hat{x}_2}^2  \Delta_{\hat{p}_2}^2  = \frac{A_R B_R + C_I^2}{4 (A_R B_R - C_R^2)}.
\end{equation}

Finally, we choose the state for the system simply as $\phi(x_3) = \exp(-x_3^2 / 2) / \pi^{1/4}$, which gives  $\Delta_{\hat{x}_3}^2 = \Delta_{\hat{p}_3}^2 = 1 / 2$.
Thus the lower bound of the generalized AK inequality given in Eq. (\ref{final-1}) is 
\begin{equation}
\label{example-6}
\Gamma = \frac{1}{2} \left[ 1 + \sqrt{ \frac{A_R B_R + C_I^2}{A_R B_R - C_R^2}} \right].
\end{equation}
If  $C_R = C_I = 0$, $\psi(x_1, x_2)$ becomes separable and hence, the original inequality (\ref{Ar-K-1}) is reproduced.  
For the entangled case  one can compute $\Delta_{\hat{x}_1} (t = 1 / \kappa) \Delta_{\hat{x}_2} (t = 1/ \kappa) = \Gamma_C$ where
\begin{equation}
\label{example-7}
\Gamma_C = \frac{\sqrt{f_1 f_2}}{8 \sqrt{A_R B_R} (A_R B_R - C_R^2)}
\end{equation}
with
\begin{eqnarray}
\label{example-8}
&&f_1 = 4 A_R B_R + (A_R B_R - C_R^2) (A_R B_R + 4 A_R + C_I^2 + 4 C_I)        \\    \nonumber
&&f_2 = 4 A_R B_R + (A_R B_R - C_R^2) (A_R B_R + 4 B_R + C_I^2 - 4 C_I).   
\end{eqnarray}
If $C_R = C_I = 0$ and $A_R = B_R = 2$, $\Gamma_C$ also becomes one.

We compare $\Gamma$ and $\Gamma_C$ in Fig. 1 when (a) $B_R = C_R = 1$ and (b) $B_R = 1$ and $C_R = 2$, The red boundary regions in both figures represent the region of $\Gamma_C \leq \Gamma$ in the $A_R - C_I$ plane.  
In this sense the correlated probe states yields the violation of the generalized AK inequality (\ref{final-1}). However, we cannot find the violation of the original AK inequality (\ref{Ar-K-1}) in this example.

\section{Conclusions}
It is needless to say that the UP is essential in the theoretical ground of quantum physics. Furthermore. it is also important from the aspect of quantum technology, because the reduction of the UP enhance the accuracy. For example, let us consider a quantum metrology\cite{metrology1,metrology2,metrology3}, which measures unknown parameters of a physical system using quantum-mechanical resources. If the parameters we want to measure are a specific pair of 
parameters associated with complementary observables, the AK inequality can yield an obstacle for the accurate measurement of the observables. In this case 
we need to find a way to circumvent the AK-inequality to increase the accuracy. The explicit protocol was examined in Ref.\cite{shapiro17}, where the
entangled-enhanced lidars are studied. The purpose of Ref.\cite{shapiro17} is to measure both a target's range and its radial velocity simultaneously. Thus, the 
AK inequality $\Delta t \Delta \omega \geq 1$ generates a significant obstacle, where $t$ and $\omega$ are the time delay and the Doppler shift. The AK inequality
can be circumvented in this reference by introducing an entanglement between transmitted signal and retained idler photons.
In this reason our main result is important, because the entangled detector can violate the generalized AK inequality in the joint measurement of the complementary variables as shown explicitly in the special example. 
Similarly, the violation of the Heisenberg sequential noise-disturbance principle was debated in Ref. \cite{lorenzo13,bullock14}.

We are not sure whether  the entangled probe states can violate the original AK inequality or not even though we have not found it in the special example. 
One can also try to find a violation of the original AK inequality by considering the tripartite entanglement\cite{tangle1}  including the state of the physical system. 
At the present stage we do not know whether the violation of the original AK inequality is possible or not. We hope to explore this issue in the future.

{\bf Data Availability}:
 The datasets generated during and/or analyzed during the current study are available from the corresponding author on reasonable request.
 
 \vspace{1.0cm}

{\bf Conflict of Interest}: 
The authors declare that they have no known competing financial interests or personal relationships that could have appeared to influence the work reported in this paper.


\newpage 

\begin{appendix}{\centerline{\bf Appendix A: Derivation of Eq. (\ref{g-AK-7}) and Eq. (\ref{g-AK-9})}}

\setcounter{equation}{0}
\renewcommand{\theequation}{A.\arabic{equation}}
In order to prove Eq. (\ref{g-AK-7})  we compute
\begin{equation}
\label{app-B-1}
J = \int dq_1 dq_2 dq_3 \delta(\theta_1) \delta (\theta_2) \delta(\theta_3) g(q_1, q_2, q_3)
\end{equation}
for arbitrary  function $g(q_1, q_2, q_3)$, First we note $ \delta(\theta_1) \delta (\theta_2) \delta(\theta_3)$ can be written as  $ \delta(X - A) \delta (Y - B) \delta (Z - C)$, where
\begin{eqnarray} 
\label{app-B-2}
&&X = q_1 + \frac{q_3}{2}   \hspace{.8cm}  Y = \frac{q_2}{m_3 \kappa} - q_3   \hspace{.8cm} Z = \frac{6 m_1 m_3 \kappa}{a} q_1 - q_2 + \frac{3 m_1 m_3 \kappa}{a} q_3       \\   \nonumber
&&A= q_1' + \frac{q_3'}{2}   \hspace{.8cm}  B = \frac{q_2'}{m_3 \kappa} - q_3'   \hspace{.8cm} C = \frac{6 m_1 m_3 \kappa}{a} q_1' - q_2' + \frac{3 m_1 m_3 \kappa}{a} q_3'.
\end{eqnarray}
Inverting Eq. (\ref{app-B-2}) one can show
\begin{equation}
\label{app-B-3}
q_1 = \left(1 - \frac{3 m_1}{a} \right) X + \frac{Y}{2} + \frac{Z}{2 m_3 \kappa}   \hspace{.5cm} q_2 = \frac{6 m_1 m_3 \kappa}{a} X - Z  \hspace{.5cm} q_3 = \frac{6 m_1}{a} X - Y - \frac{Z}{m_3 \kappa}.
\end{equation}
Computing the Jacobian, one can show $dq_1 dq_2 dq_3 = dX dY dZ$. 
Thus, $J$ can be written as $J = g(q_1', q_2', q_3')$, which implies  Eq. (\ref{g-AK-7}). 

Eq. (\ref{g-AK-9}) can be derived by making use of Eq. (\ref{app-B-3}) as follows:
\begin{equation}
\label{app-B-4}
\frac{\partial}{\partial \theta_1} \equiv \frac{\partial}{\partial X} = \frac{\partial q_1}{\partial X} \frac{\partial}{\partial q_1} + \frac{\partial q_2}{\partial X} \frac{\partial}{\partial q_2} + \frac{\partial q_3}{\partial X} \frac{\partial}{\partial q_3}
= \left(1 - \frac{3 m_1}{a} \right) \frac{\partial}{\partial q_1} + \frac{6 m_1 m_3 \kappa}{a} \frac{\partial}{\partial q_2} + \frac{6 m_1}{a} \frac{\partial}{\partial q_3}.
\end{equation}
Similarly, it is easy to show
\begin{equation}
\label{app-B-5}
\frac{\partial}{\partial \theta_2} = \frac{1}{2} \frac{\partial}{\partial q_1} - \frac{\partial}{\partial q_3}.
\end{equation}
Then, taking a partial integration one can prove Eq. (\ref{g-AK-9}). Furthermore, it is easy to prove Eq. (\ref{g-AK-13}) by taking partial integration twice.

\end{appendix}

\end{document}